\documentclass[reprint,nofootinbib,amsmath,amssymb,prl]{revtex4-2}

\usepackage{graphicx}
\usepackage{dcolumn}
\usepackage{bm}
\usepackage{xcolor}
\usepackage{wrapfig}
\usepackage{multirow}

\renewcommand{\vec}[1]{\boldsymbol{#1}}
\newcommand{\be}{\begin{equation}}
\newcommand{\ee}{\end{equation}}
\newcommand{\bea}{\begin{eqnarray}}
\newcommand{\eea}{\end{eqnarray}}
\newcommand{\Tr}{\,\hbox{\rm Tr}}

\def\MSbar{\overline{\rm MS\kern-0.5pt}\kern0.5pt}
\def\psibar{\overline{\psi}}

\begin{document}

\title{QCD Equation of State with $N_f=3$ Flavors up to
the Electroweak Scale}

\author{Matteo Bresciani$^{a,b}$}
\author{Mattia Dalla Brida$^{a,b}$}
\author{Leonardo Giusti$^{a,b}$}
\author{Michele Pepe$^{b}$}
\affiliation{
$^a$ Dipartimento di Fisica, Universit\`a di Milano-Bicocca, Piazza della Scienza 3, 
I-20126 Milano, Italy\\
$^b$ INFN, Sezione di Milano-Bicocca, Piazza della Scienza 3, 
I-20126 Milano, Italy}
                
\date{\today}

\begin{abstract}
The equation of state of Quantum Chromodynamics with $N_f=3$ flavors
is determined non-perturbatively in the range of temperatures between $3$ and $165$~GeV
with a precision of about $0.5$-$1.0$\%. The calculation is
carried out by numerical simulations of lattice gauge theory discretized \`a la Wilson
with shifted boundary conditions in the compact direction. At each given temperature the entropy
density is computed at several lattice spacings in order to extrapolate
the results to the continuum limit. Taken at face value,
data point straight to the Stefan-Boltzmann value by following a linear
behavior in the strong coupling constant squared. They are also compatible
with the known perturbative formula supplemented by higher order terms
in the coupling constant, a parametrization which describes well our data
together with those present in the literature down to $500$~MeV. 
\end{abstract}

\maketitle

\textbf{\textit{Introduction}} ---
The Equation of State (EOS) of Quantum Chromodynamics (QCD) describes the collective
behavior of strongly-interacting particles at equilibrium. It is of absolute interest in
particle and nuclear physics, and in cosmology where it plays a crucial role in many
physical processes. It has contributed to the evolution
of the effective numbers of degrees of freedom in the thermal expansion of the
Universe, see Refs.~\cite{ParticleDataGroup:2024cfk,Saikawa:2018rcs} for
a recent discussion and references therein. The evolution during the
QCD epoch may have left distinctive features in the
spectrum of the primordial gravitational waves background~\cite{Saikawa:2018rcs}, or may have
affected significantly the abundance of dark matter candidates, e.g.
WIMPs~\cite{Hindmarsh:2005ix,Drees:2015exa,Saikawa:2020swg} and/or axions
~\cite{Borsanyi:2016ksw,GrillidiCortona:2015jxo}. In terrestrial
laboratories, QCD matter under extreme conditions is produced and investigated
at heavy-ion colliders, where the EOS is an essential input in the analysis of data.\\
\indent At present, the EOS of QCD at zero chemical potential is known up to
temperatures of about $1$~GeV and
with a precision of a few percent. It has been computed non-perturbatively
by extrapolating to the continuum limit lattice QCD results
with $N_f=2+1$ \cite{Borsanyi:2013bia,HotQCD:2014kol,Bazavov:2017dsy} and $N_f=2+1+1$
\cite{Borsanyi:2016ksw} flavors. Above $1$~GeV, the EOS is known
only perturbatively as a function of the strong coupling constant
$g$. Its expansion has been computed up to order $g^6 \log(g)$,
after which the non-perturbative terms from the ultrasoft modes
start to contribute
~\cite{Linde:1980ts,Kajantie:2002wa}, see Ref.~\cite{Navarrete:2024ruu} for recent
progress in this field. The convergence of the expansion, however,
turns out to be very poor, and  the perturbative
expression remains insufficient to achieve reliable and precise cosmological
predictions~\cite{Saikawa:2018rcs}. On the theoretical side, the calculation of
the EOS in the SU($3$) Yang--Mills theory showed that the
contributions computed in perturbation theory are very far from
explaining the full result even at temperatures as high as the electroweak
scale~\cite{Borsanyi:2012ve,Giusti:2016iqr}. More recently, the non-perturbative
computation of the QCD screening masses showed that the known perturbative results
do not provide a satisfactory description up to these high
temperatures~\cite{DallaBrida:2021ddx,Giusti:2024ohu}. All these results indicate
that, for a reliable and precise determination of its thermal properties,
a fully non-perturbative treatment of QCD is required up to temperatures
much higher than $1$~GeV.\\
\indent The purpose of this Letter is to compute the EOS of QCD non-perturbatively
over a wide range of
temperatures $T$ from $3$~GeV up to the electroweak scale.
The computational strategy is entirely new, and we focus on the theory
with $N_f=3$ flavors of massless quarks.
This choice is justified since the effect of the masses of the $3$ lightest quarks is
negligible within errors in the explored range of temperatures~\cite{Laine:2006cp,Borsanyi:2016ksw,Bazavov:2017dsy}.
At the same time, the strategy adopted is fully general
and can be applied to QCD with five (massive) quarks without further conceptual difficulties.\\
\indent This progress is achieved by extending and applying to the EOS a recently proposed strategy
for studying QCD up to very high temperatures from first
principles~\cite{DallaBrida:2021ddx}.  The key novelties with respect to the
standard ``integral'' method are: (a) abandon the hadronic scheme
to renormalize the theory, and determine the lines of constant physics by fixing the
value of a non-perturbatively defined renormalized running coupling computed at very different scales
by using step-scaling techniques~\cite{Luscher:1991wu,Luscher:1993gh,Jansen:1995ck},
(b) avoid the zero-temperature subtraction of the ultraviolet power divergence in
thermodynamic potentials by computing directly
the entropy density $s$ in a moving reference frame formalized by shifted boundary
conditions~\cite{Giusti:2010bb,Giusti:2011kt,Giusti:2012yj,DallaBrida:2020gux}.
The pressure $p$ can then be computed by integrating the entropy in the temperature, while the energy density
$e$ is determined by integrating the temperature in the entropy or, equivalently, by using
the relation $T s =(e+p)$.\\[0.25cm]
\noindent \textbf{\textit{Preliminaries}} ---
We regularize QCD on a finite four-dimensional lattice of spatial volume $V=L^3$,
compact extension $L_0$, and spacing $a$.
The fields satisfy periodic boundary conditions in the 
spatial directions and shifted boundary conditions in the compact one 
\bea
   U_\mu(x_0+L_0,\vec x)  & = & U_\mu(x_0,\vec x - L_0\vec\xi)\; ,\nonumber\\
   \psi(x_0+L_0,\vec x)  & = &  -\psi(x_0,\vec x - L_0\vec\xi)\; ,\label{eq:shift}\\
\psibar(x_0+L_0,\vec x)  & = & -\psibar(x_0,\vec x - L_0\vec\xi)\;,\nonumber
\eea
where $U_\mu\in$ SU($3$) are the link variables, $\psi$ and
$\psibar$ are triplets in flavor space of degenerate quark fields, and the
spatial vector $\vec \xi$ characterizes the moving frame in the Euclidean
space-time~\cite{Giusti:2010bb,Giusti:2011kt,Giusti:2012yj}.\\ 
\indent The QCD lattice action is $S=S_G+S_F$, where $S_G$ and $S_F$ are the gluonic 
and the fermionic parts, respectively. The pure gauge action is discretized
through the standard Wilson plaquette definition
\be\label{eq:latS}
S_G = \frac{6}{g_0^2}\, \sum_{x} \sum_{\mu<\nu} 
\left[1 - \frac{1}{3}{\rm Re}\Tr\Big\{U_{\mu\nu}(x)\Big\}\right]\; ,
\ee
where the trace is over the color index, $g_0$ is the bare coupling
constant, $U_{\mu\nu}(x)$ is the plaquette as defined in
Ref.~\cite{DallaBrida:2021ddx}, $\mu,\,\nu=0,\dots,3$,
and $x$ is the space-time coordinate. The fermionic part of the
action is 
\begin{equation}\label{eq:SF}
S_F=a^4\sum_x \psibar(x) (D+m_0)\psi(x)\; ,
\end{equation}
where $m_0$ is the bare quark mass and the $O(a)$-improved
lattice Dirac operator is
$D=D_{\rm w} + a D_{\rm sw}$~\cite{Wilson:1974sk,Sheikholeslami:1985ij}.
The massless Wilson-Dirac operator $D_{\rm w}$
and the Sheikholeslami-Wohlert operator $D_{\rm sw}$ 
are given in Ref.~\cite{DallaBrida:2021ddx}
where all unexplained notation adopted in
this Letter can be found. Finally, the partition function $Z_{\vec\xi}$
and the free energy density $f_{\vec\xi}$ are readily
defined as
\be
f_{\vec\xi}= -\frac{1}{L_0 V}\ln{Z_{\vec\xi}}\;, \qquad Z_{\vec\xi} = \int DU D\psi D\psibar\; e^{-S}\; ,
\ee
where the integration measures on the various fields are
defined as usual.

The key idea for reaching very high temperatures with a moderate computational
effort is to renormalize the theory by fixing the value of a non-perturbatively defined
coupling~\cite{Giusti:2016iqr,DallaBrida:2021ddx}. The coupling can
be defined in a finite volume and computed precisely and efficiently on the lattice for values of
the renormalization scale $\mu$ which span several orders of
magnitude by using step-scaling techniques~\cite{Luscher:1993gh,Brida:2016flw,DallaBrida:2018rfy}.
To make a definite choice,
we adopt the definition based on the Schr\"odinger functional
(SF) $\bar g^2_{\rm SF}(\mu)$~\cite{Luscher:1992an,Sint:1993un,Sint:1995ch}. Once the renormalized coupling
has been determined in the continuum limit for $\mu \sim T$ \cite{Brida:2016flw,DallaBrida:2018rfy},
the theory at temperature $T$ is renormalized by fixing the coupling at
finite lattice spacing to be
\be
\bar g^2_{\rm SF}(g_0^2, a\mu) = \bar g^2_{\rm SF}(\mu)\; ,\quad  a\mu\ll 1\;.
\ee
This condition fixes the so-called lines of constant physics, i.e.
the dependence of the bare coupling constant $g_0^2$ on the lattice spacing, for
values of $a$ at which the scale $\mu$ and therefore $T$ can be
easily accommodated. As we change the physical temperature,
we impose different
renormalization conditions which, however, define the very same renormalized
theory at all temperatures.
Once the lines of constant physics have been fixed, we determine the values of
the critical mass $m_{\rm cr}$ by requiring that the PCAC mass,
computed in a finite volume with SF boundary conditions, vanishes~\cite{DallaBrida:2018rfy}. As
a consequence, at each $T$ the theory can be simulated efficiently at various
lattice spacings without suffering from large discretization errors, and the continuum limit
of the observables can be taken with confidence. All technical details on how the
renormalization procedure is implemented in practice and the values of the bare parameters of the
lattices simulated are given in Appendices A and B of
Ref. \cite{DallaBrida:2021ddx}.\\[0.25cm]
\indent \textbf{\textit{Entropy density in a moving frame}} ---
In the presence of shifted boundary conditions, the entropy density at finite
lattice spacing can be defined as~\cite{Giusti:2010bb,Giusti:2011kt,Giusti:2012yj}
\be\label{eq:sshift}
\frac{s}{T^3} = \frac{1+\vec\xi^2}{\xi_k} \frac{1}{T^4} \frac{\Delta f_{\vec\xi}}{\Delta \xi_k}
\ee
where $T=1/(L_0 \sqrt{1+\vec\xi^2})$, and the last term on the r.h.s.~is a discrete derivative
of the free-energy density with respect to the shift component in direction $k$. For
computational efficiency, it is convenient to introduce the following decomposition
at fixed $g_0$ and $L_0/a$ 
\be\label{eq:deltaf}
\frac{\Delta f_{\vec\xi}}{\Delta \xi_k} = 
\frac{\Delta
  (f_{\vec\xi}-f^\infty_{\vec\xi})}{\Delta \xi_k} +
\frac{\Delta f^\infty_{\vec\xi}}{\Delta \xi_k}\;,\\[0.125cm]
\ee
where $f^\infty_{\vec\xi}$ is the free-energy density in the limit $m_q\rightarrow\infty$, and
$m_q=m_0-m_{\rm cr}$ is the subtracted quark mass. By rewriting 
the first term on the r.h.s.~of Eq.~(\ref{eq:deltaf}) as the integral of the
derivative w.r.t.~$m_q$, we have
\be
\frac{\Delta(f_{\vec\xi}\!-\! f^\infty_{\vec\xi})}{\Delta \xi_k} = 
- \frac{\Delta}{\Delta \xi_k} \int_{m_q}\!\!\!
\frac{\partial f^{m_q}_{\vec\xi}}{\partial m_q} = 
- \int_{m_q}\!\!\!
\frac{\Delta \langle \psibar \psi \rangle^{m_q}_{\vec\xi}}{\Delta \xi_k}\; ,
\label{eq:deltafm}
\ee
where $\int_{m_q}=\int_{0}^{\infty} d m_q$, and 
the last integral can be computed by performing numerical simulations of QCD at
various quark masses, as we discuss below.
The second term on the r.h.s.~of
Eq.~(\ref{eq:deltaf}) corresponds to the contribution in the static quark limit
of QCD, i.e. in the SU($3$) Yang--Mills theory, and it can be
written as~\cite{Giusti:2015daa} 
\be\label{eq:deltafg}
\frac{\Delta f^\infty_{\vec\xi}}{\Delta \xi_k} = \frac{\Delta f^\infty_{\vec\xi}}{\Delta \xi_k}\Bigg|_{g_0^2=0}
- \int_{0}^{g_0^2} d u\,\frac{1}{u} \frac{\Delta \langle S_G \rangle^\infty_{\vec\xi}}{\Delta \xi_k}\Bigg|_{g_0^2=u}\; .
\ee
Although this contribution has the largest variance, it can be computed by pure
gauge simulations which are orders of magnitude cheaper than those of QCD.\\[0.25cm]
\begin{figure*}[th]
  \vspace{1.5cm}
\begin{center}
\begin{minipage}{0.48\textwidth}
\includegraphics[width=8.5 cm,angle=0]{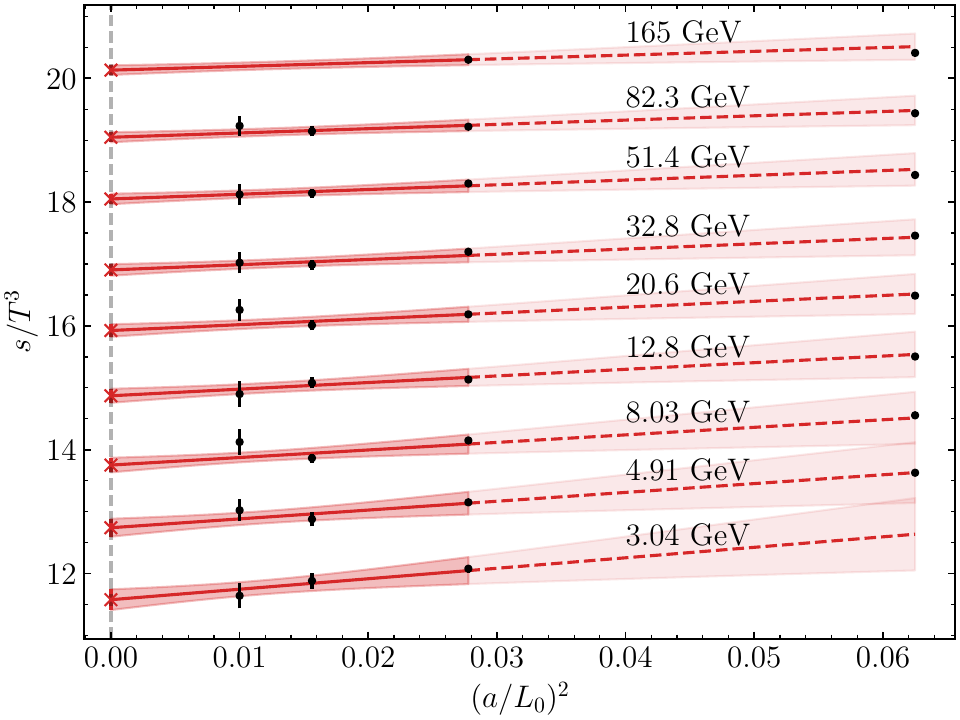}
\end{minipage}
\hspace{5mm}
\begin{minipage}{0.48\textwidth}
\includegraphics[width=8.5 cm,angle=0]{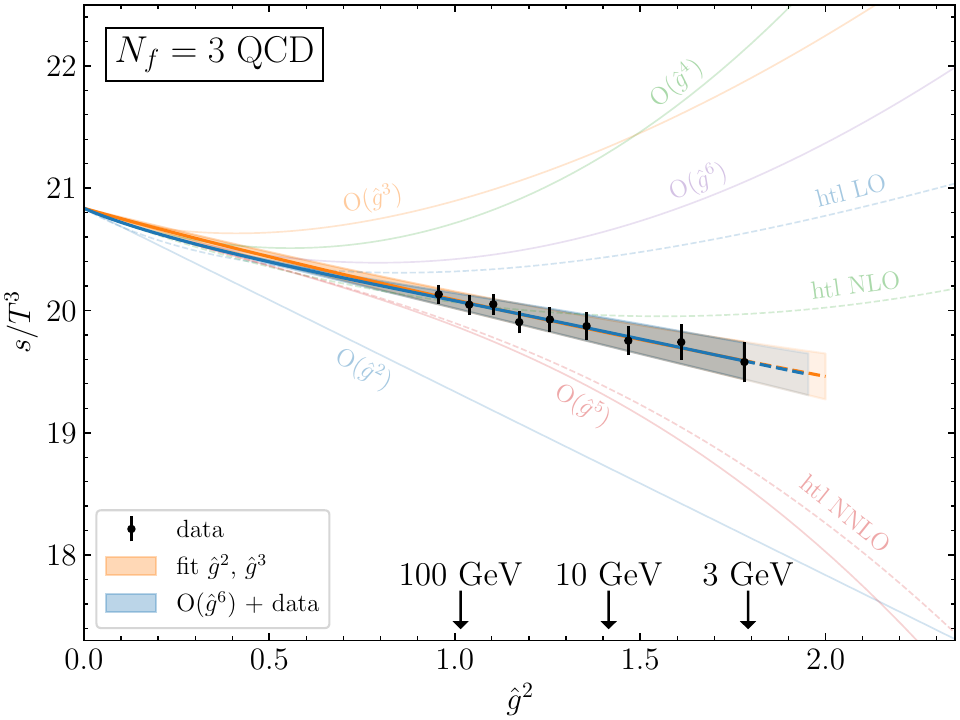}
\end{minipage}
\caption{Left: values of $s/T^3$ as a function of $(a/L_0)^2$ corresponding to $T_n$ ($n=0,\dots,8$) shifted
downward by $n$ for better readability. Right: continuum values of $s/T^3$ versus
$\hat g^2(T)$. The various continuous curves show the central values of the perturbative expressions
in Ref.~\cite{Kajantie:2002wa}, each one including up to the order indicated on the curve. The analogous
curves for hard-thermal-loop (htl) perturbation theory~\cite{Andersen:2011sf} are represented by dashed lines.
The orange and the blue bands (gray when the two overlap) represent the results of
the fits to Eqs.~(\ref{eq:ints}) and (\ref{eq:LatpPT}), respectively,
as explained in the main text.
\label{fig:results}}
\end{center}
\end{figure*}
\textbf{\textit{Numerical computation of $s$}} ---
We compute the entropy density in QCD with $N_f=3$ massless
quarks at the $9$ temperatures $T_0,\dots,T_{8}$ given in Table~\ref{tab:s_CL}
which span the range from about $3$ GeV up to $165$ GeV. We adopt shifted boundary
conditions in the compact direction with $\vec \xi=(1,0,0)$
and, in order to extrapolate the results to the continuum limit with confidence, several
lattice spacings are simulated at each temperature with the extension of the compact direction
being $L_0/a=4,6,8,10$.
\begin{wraptable}{r}{3.5cm}
\begin{tabular}{|c|c|c|}
\hline
 & & \\[-0.25cm]
$T$ & $T({\rm GeV})$  & $s/T^3$\\[-0.25cm]
  & & \\
\hline
$T_0$ &  165(6)    & 20.13(8) \\
$T_1$ &  82.3(2.8) & 20.05(8)  \\
$T_2$ &  51.4(1.7) & 20.05(9) \\
$T_3$ &  32.8(1.0) & 19.90(9) \\
$T_4$ &  20.6(6)   & 19.93(10) \\
$T_5$ &  12.8(4)   & 19.87(11) \\
$T_6$ &  8.03(22)  & 19.75(12)  \\
$T_7$ &  4.91(13)  & 19.74(15)  \\
$T_8$ &  3.04(8)   & 19.58(17)  \\
\hline
\end{tabular}
\caption{Temperatures considered together with the results for
  $s/T^3$ in the continuum limit.\label{tab:s_CL}}
\end{wraptable} 
The inverse bare coupling constant $6/g_0^2$ and the critical quark
mass are fixed at each lattice spacing by applying the renormalization strategy outlined above.
The values of $6/g_0^2$ simulated are reported in Table~\ref{tab:matel} of Appendix~C; the full set of bare parameters is
given in Table~4 of Ref. \cite{DallaBrida:2021ddx}. The lattice
size in the spatial directions has been set to $L/a=144$ so that $L T$ ranges from
$10$ to $25$. We have explicitly checked that finite size effects in the
entropy density are negligible within our statistical errors, see Appendix~D for details.
The discrete derivatives in the
shift are estimated by computing the chiral condensate
and the expectation values of the gluon action at ${\vec \xi}=(1\pm 2a/L_0,0,0)$, so that each
evaluation requires $2$ different simulations.
For each value of $L_0/a$ and $6/g_0^2$, the integral in Eq.~(\ref{eq:deltafm})
is carried out by combining Gaussian quadratures as described in Appendix~A which, altogether, require
the computation of the derivative of the chiral condensate at $20$
quark masses. The integral on the r.h.s.~of Eq.~(\ref{eq:deltafg}) is computed from the 
the discrete derivative of the numerical estimate of $\langle S_G \rangle^\infty_{\vec\xi}$
in the SU($3$) Yang--Mills theory as discussed in Appendix~B.\\
\begin{figure*}[th]
  \vspace{1.5cm}
\begin{center}
\begin{minipage}{0.48\textwidth}
\includegraphics[width=8.5 cm,angle=0]{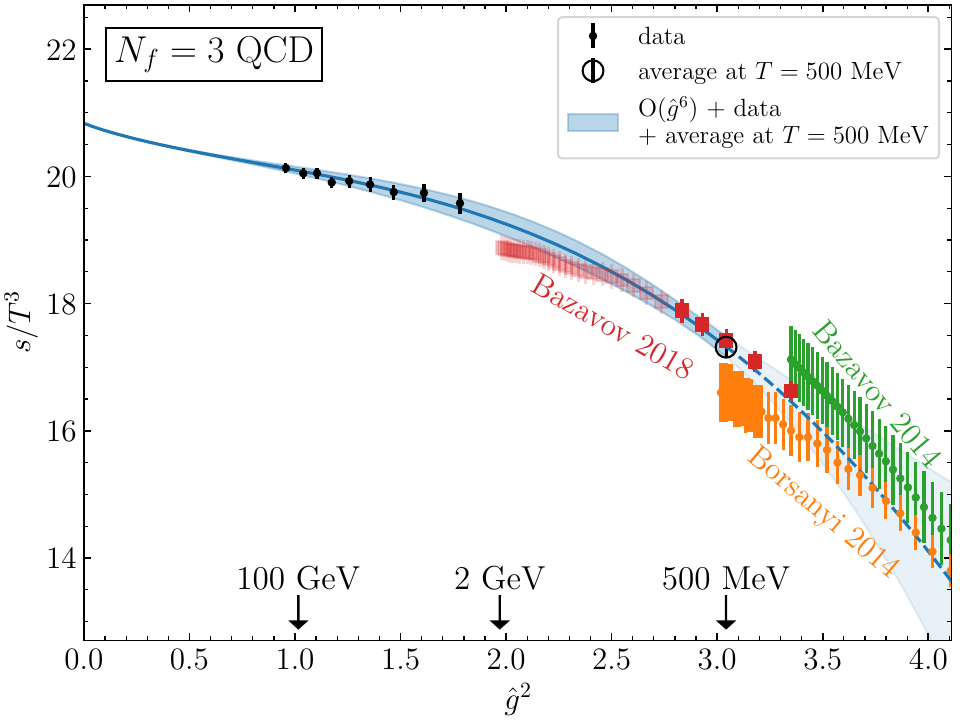}
\end{minipage}
\hspace{5mm}
\begin{minipage}{0.48\textwidth}
\includegraphics[width=8.5 cm,angle=0]{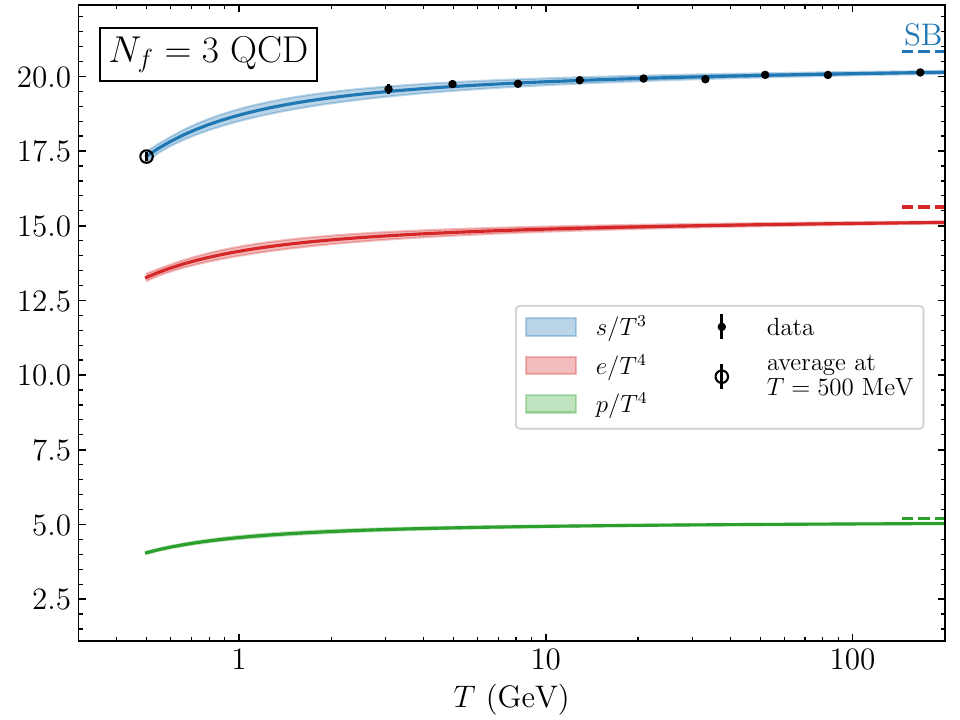}
\end{minipage}
\caption{Left: normalized entropy density, $s/T^3$, versus $\hat g^2(T)$. The blue curve is our
  best parametrization of $s/T^3$ for $T\geq 500$~MeV. Data from
  Refs.~\cite{Borsanyi:2013bia,HotQCD:2014kol,Bazavov:2017dsy} are also shown.
  Right: normalized pressure, entropy and energy densities as a function of temperature
  for $T\geq 500$~MeV. \label{fig:bestresult}}
\end{center}
\end{figure*}
\indent  In the left panel of Fig.~\ref{fig:results} we show the one-loop perturbatively
improved values, as defined in Appendix~E, of $s/T^3$. To extrapolate these results to the continuum
limit, we have considered several fitting strategies which are discussed in detail in Appendix~F.
The best fit is the one where points at all temperatures and $L_0/a>4$ are fitted
together linearly in $(a/L_0)^2$ with a coefficient
whose temperature dependence is proportional to $\bar g^3_{\rm SF}(\sqrt{2}\, T)$ and
with weights in the definition of the $\chi^2$ which take into account statistical
and systematic errors. The resulting value of $\chi^2/\chi^2_{\rm exp}$ is $0.82$,
where $\chi^2_{\rm exp}$ is defined as in Ref.~\cite{Bruno:2022mfy}.
Lattice artifacts turn out to be rather mild.\\
\indent Our best values of $s/T^3$ in the continuum limit are reported
in Table~\ref{tab:s_CL} and shown in the right panel of Fig.~\ref{fig:results}
as a function of $\hat g^2(T)$. The latter is defined as the five-loop
strong coupling constant squared in the $\MSbar$ scheme at the renormalization scale $\mu=2\pi T$.
At the leading order, its well known expression reads
\begin{equation}\label{eq:gmu}
  \frac{1}{\hat g^2(T)} \equiv \frac{9}{8\pi^2} \ln  \frac{2\pi T}{\Lambda_{\MSbar}} +\ldots \; , 
\end{equation}
where $\Lambda_{\MSbar} = 341$~MeV is taken from Ref.~\cite{Bruno:2017gxd}, while
for the complete formula see Ref.~\cite{Baikov:2016tgj,inprep}. It is important
to stress that for our purposes $\hat g^2(T)$ is just a function of the temperature $T$
that we use to scrutinize in detail the temperature dependence of our results. Taken
at face value, the data point straight to the Stefan-Boltzmann (SB) value from which the point at
the highest temperature differs by approximatively $3\%$.
The entropy density is indeed well represented by the phenomenological formula
\be\label{eq:ints}
\frac{s}{T^3} = \frac{32\pi^2}{45}
\left[s_0 + s_2 \, \Big(\frac{\hat g}{2\pi}\Big)^2 + s_3 \Big(\frac{\hat g}{2\pi}\Big)^3 \right]\; 
\ee
over the almost 2 orders of magnitude of the temperature range explored.
If one opts for the simplest choice by forcing $s_3=0$, then the fit
coefficients are $s_0= 2.954(15)$ and $s_2= -3.6(7)$ with the off diagonal elements of
the covariant matrix being ${\rm cov}(s_0,s_2)/[\sigma(s_0)\sigma(s_2)]=-0.84$. This result confirms that,
when extrapolated to the $T\to\infty$ limit, the data point to the
SB value, $s^{\rm SB}_0=2.969$, within errors. By enforcing $s_0=s^{\rm SB}_0$, the
fit parameters are $s_2= -5.1(9)$, $s_3= 5(5)$ and the off diagonal
element of the covariant matrix is ${\rm cov}(s_2,s_3)/[\sigma(s_2)\sigma(s_3)]=-0.89$.
The corresponding curve is shown in orange in the right panel of Fig.~\ref{fig:results}.
These analyses show that our data are compatible
with a linear behavior in $\hat g^2(T)$ but with a slope which tends to be
different from the perturbative value $-8.438$.\\[0.25cm]
\textbf{\textit{The Equation of State in the continuum}} ---
The known perturbative formulas suggest to parametrize
the non-perturbative data with a polynomial interpolation of the form
\be\label{eq:LatpPT}
\frac{s}{T^3} = \frac{32\pi^2}{45} \sum_{k} s_k \left(\frac{\hat g}{2\pi}\right)^{k}\; .
\ee
By fixing the first coefficients to their perturbative values,   
$s_0=2.969$, $s_1=0$, $s_2=-8.438$, $s_3=55.11$, $s_4=-40.28+101.2\, \ln(\hat g^2)$,
$s_5=-1174$, $s_6=4791-1629\, \ln(\hat g^2)+q_c$ guarantees the expected
behavior of the normalized entropy density at asymptotically
large temperatures~\cite{Kajantie:2002wa}. As usual, we introduce the free coefficient $q_c$
because there are unknown contributions at this order including those of non-perturbative
origin in the three-dimensional effective theory.
The last term that we include in the polynomial interpolation 
is of $O(\hat g^7)$.
By fitting our data to the functional form
in Eq.~(\ref{eq:LatpPT}), we obtain $q_c=-5.1(1.7)\cdot 10^3$, $s_7=1.3(7)\cdot 10^4$
with ${\rm cov}(q_c,s_7)/[\sigma(q_c)\sigma(s_7)]=-0.99$ and
an excellent $\chi^2/\chi^2_{\rm exp}=0.58$. The corresponding curve
is shown in blue in the right panel of Fig.~\ref{fig:results}
(gray when this curve overlaps with the orange one).
If we force $s_7=0$, we obtain $q_c=-2.5(3)\cdot 10^3$ with
$\chi^2/\chi^2_{\rm exp}=1.52$. This analysis shows that contributions
beyond the known perturbative ones are required to explain
data even at these very large temperatures. For completeness it
must be said that the data are compatible with the results of hard-thermal-loop
perturbation theory, albeit
within the large systematic uncertainties quoted in Ref.~\cite{Andersen:2011sf}.
A detailed discussion
of this comparison will be reported in Ref.~\cite{inprep}.\\[0.25cm]
\textbf{\textit{Discussion and conclusions}} ---
At much lower temperatures, up to $500$~MeV or so, 
the entropy density with $N_f=2+1$ flavors has been
determined in Refs.~\cite{Borsanyi:2013bia,HotQCD:2014kol,Bazavov:2017dsy}
with percent precision by extrapolating lattice
data to the continuum limit. It is appropriate to compare
their results with ours because the contribution due to the mass
of the light quarks is at the permille level at
$500$~MeV~\cite{Laine:2006cp}, well below the errors quoted
by the two collaborations. Albeit with very large errors,
the best fit of our data in the previous section is
compatible with the value of the normalized entropy density
at $500$~MeV, $s/T^3=17.31(16)$, obtained by averaging the two results
from Refs.~\cite{Borsanyi:2013bia,Bazavov:2017dsy}. We therefore
combine the latter value with our data, and interpolate
this enlarged data set with the polynomial given in
Eq.~(\ref{eq:LatpPT}). Again the first coefficients are fixed
to their perturbative values, while a fit to the data gives
$q_c = -4.0(1.1)\cdot 10^3$, $s_7=7(4)\cdot 10^3$ with
the off diagonal element of the covariant matrix being
${\rm cov}(q_c,s_7)/[\sigma(q_c)\sigma(s_7)]=-1.00$ and
$\chi^2/\chi^2_{\rm exp}=0.79$. This is our best parametrization of
$s/T^3$ for QCD with $N_f=3$ flavors for $T\geq 500$~MeV which
has an error of at most $1$\% within that range. It 
is shown in blue in the left panel of Fig.~\ref{fig:bestresult}.
Since $s_7$ turns out to be close to zero, for
completeness we also provide the result $q_c = -2.02(6)\cdot 10^3$
of the fit obtained by fixing $s_7=0$ which has a
$\chi^2/\chi^2_{\rm exp}=2.64$. We notice that,
when extended below $500$~MeV, our best interpolation curve is
compatible with the continuum extrapolated data in
Refs.~\cite{Borsanyi:2013bia,HotQCD:2014kol,Bazavov:2017dsy} given
the large errors of the data,  see left panel of
Fig.~\ref{fig:bestresult}. \\
The pressure can be obtained by integrating $s(T)$ starting from $p(0)=0$, requiring
to compute the entropy density down to low temperatures. In this Letter we exploit the parametrization in
Eq.~(\ref{eq:LatpPT}) which was derived by applying the thermodynamic relation $s = \partial p/\partial T$
to the analogous expression for the pressure~\cite{Kajantie:2002wa}
\be\label{eq:presPT}
\frac{p}{T^4} = \frac{8\pi^2}{45} \sum_{k} p_k \left(\frac{\hat g}{2\pi}\right)^{k}\; .
\ee
The coefficients of the two expansions are, thus, related and the first coefficients for the pressure similarly
take their perturbative values: $p_0=s_0$, $p_1=s_1$, $p_2=s_2$, $p_3=s_3$, $p_4=-49.77+101.2\, \ln(\hat g^2)$, and
$p_5=-1081$. The next coefficients $p_6=4776-1401\, \ln(\hat g^2)+q_c$ and
$p_7=s_7 + (5\, b_0 p_5 + 3\, b_1 p_3)/4$, with $b_0=9/4$, $b_1=4$, contain non-perturbative contributions
that are included in the parameters $q_c$ and $s_7$ fitted above. The energy density is obtained as
$e = T s - p$. The corresponding curves are shown in the right panel of Fig.~\ref{fig:bestresult} for $T\geq 500$~MeV.\\
The strategy introduced here has paved the way to determine the EOS
of QCD up to temperatures of the order of the electroweak scale.
The numerical results presented here can indeed be systematically
improved in the future by investing more computational resources.
As a final remark, we notice that the strategy can be
straightforwardly extended to QCD with 4 and 5 (massive)
quarks.\\[0.25cm]
\textbf{\textit{Acknowledgments}} ---
We  acknowledge Leonardo Cosmai for useful discussions.
We acknowledge PRACE for awarding us access to the HPC system MareNostrum4 at the Barcelona Supercomputing
Center (Proposals n. 2018194651 and 2021240051) where some of the numerical results presented in this
Letter have been obtained. We also thank CINECA for providing us with a very generous access
to Leonardo during the early phases of operations of the machine and for the computer time
allocated via the CINECA-INFN, CINECA-Bicocca agreements. The R\&D has been carried out on the PC
clusters Wilson and Knuth at Milano-Bicocca. We thank all these institutions for the technical
support. This work is (partially) supported by Italian Center for SuperComputing (ICSC) – Centro Nazionale di Ricerca in High
Performance Computing, Big Data and Quantum Computing, funded by European Union – NextGenerationEU.

\bibliography{bibfile}


\clearpage
\newpage
\begin{center}
\textbf{\textit{Appendices}}
\end{center}
 
\textbf{\textit{Appendix A: Integration in the bare mass}} ---
The numerical computation of the integral in the bare subtracted quark mass 
of Eq.~\eqref{eq:deltafm} is carried out as follows.
At fixed values of $L_0/a$ and $g_0^2$ we split the integration 
in three intervals of bare mass $\left\{0\leq m_q/T< 5\right\}$, 
$\left\{5\leq m_q/T< \widetilde{m}\right\}$ and
$\left\{\widetilde{m}\leq m_q/T<\infty\right\}$, where
$\widetilde{m}=35$ for $L_0/a=4$ and $\widetilde{m}=20$ for $L_0/a=6, 8, 10$.
The integrals in the first two domains are computed with 
10-point and 7-point Gaussian quadratures, respectively.
After a change of variable to the hopping parameter $\kappa=1/(2am_0+8)$, which 
makes the third domain compact, the integral is computed with a 3-point
Gaussian quadrature. This optimized integration scheme has been chosen using tree-level perturbation 
theory as guidance so that systematic effects from the numerical quadratures are 
negligible with respect to the target statistical accuracy. The integration
region was split such that approximately 80\% of the integral at tree-level
came from the first region and about 20\% from the second part, with the
remainder contributing at most about 1$\sigma$. The choice of 10, 7, and 3 points for the
three regions was made so that systematic effects were negligible, at most contributing by
less than $0.1\,\sigma$. The shape of the non-perturbative curve is very similar to the one
obtained from tree-level perturbation theory supporting the validity of the systematic effects analysis.
In order to ensure full confidence, we checked explicitly on the non-perturbative
data of a selected set of lattices that the systematic effects from the numerical quadratures remain
negligible by using the Gauss-Kronrod quadrature,
i.e. the result does not change within errors by adding integration points.
\begin{figure}[h]
    \centering
    \includegraphics[width=0.4\textwidth]{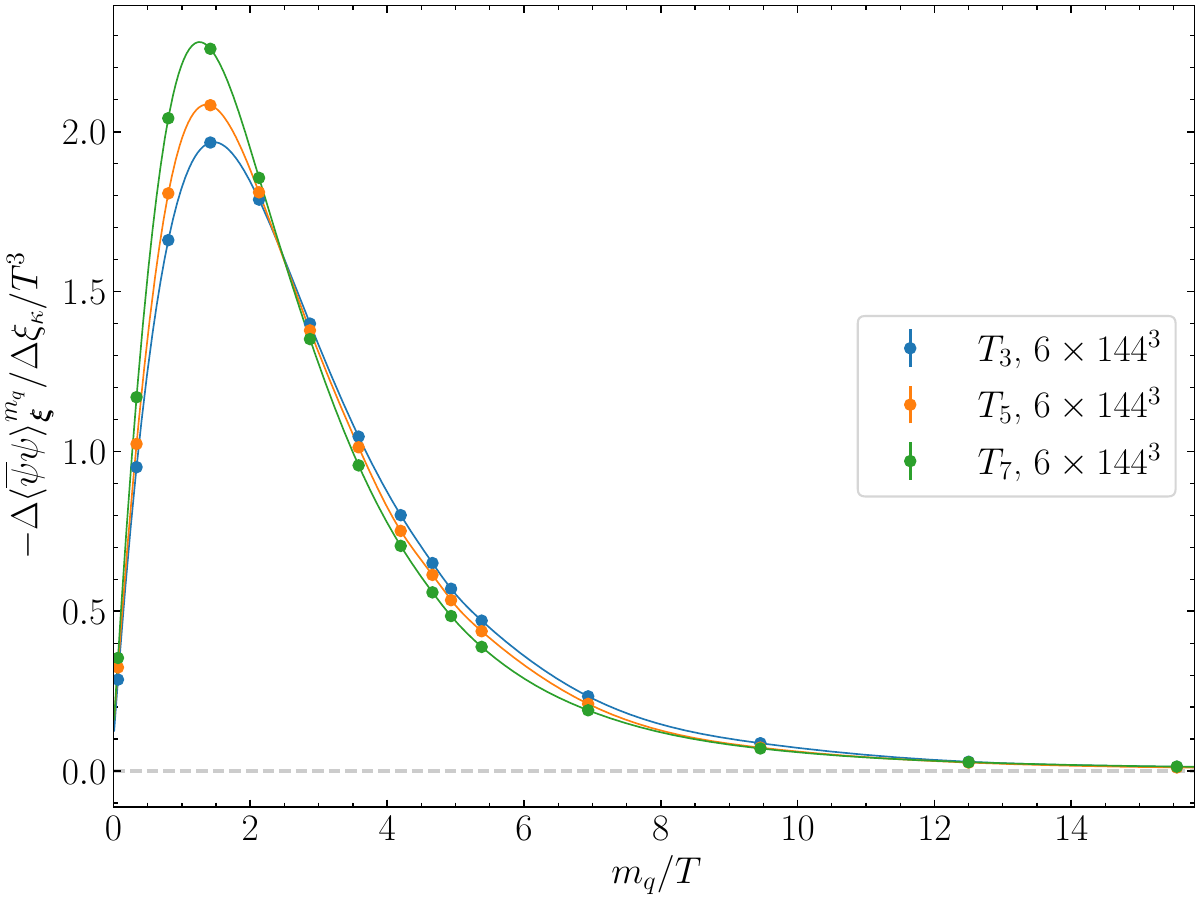}
    \caption{Derivative in the shift of the chiral condensate as a function 
    of $m_q/T$ at some selected bare parameters. Points have been interpolated with a 
    cubic spline to guide the eye. Error bars are smaller than the markers.}
    \label{fig:mass_integral}
\end{figure}

Figure~\ref{fig:mass_integral} shows the integrand function as obtained 
from lattice simulations. Each point in the plot comes from two independent
ensembles with shifts $\vec\xi =(1\pm 2a/L_0,0,0)$ at the same bare
parameters (see main text).\\[0.25cm]
\textbf{\textit{Appendix B: Integration in the bare coupling}} ---
At fixed $L_0/a$, the integration in the bare coupling appearing in
the right side of Eq.~\eqref{eq:deltafg} has to be performed 
in the intervals $g_0^2\in[0,g_0^2|_{T_i}]$, $i=0,\ldots,8$, 
where $g_0^2|_{T_i}$ are the bare couplings at given temperature $T_i$ whose corresponding
inverse values $6/g_0^2$ are reported in Table~\ref{tab:matel}.
The integrals have been split into several parts. 
For the first segment, $[0, 6/15]$, we applied the 2-point trapezoidal rule (3-point Simpson rule for $L_0/a=4$) 
and a 3-point Gaussian quadrature has been employed in the range $[6/15,6/9]$.
In the domain $[6/9, g_0^2|_{T_0}]$ we adopted a 3-point Gaussian quadrature for $L_0/a=4$, 
and the midpoint rule for $L_0/a=6$.
The interval $[6/9,g_0^2|_{T_1}]$ has been integrated with a 3-point 
Gaussian quadrature, as well as the domains
$[g_0^2|_{T_{i-1}}, g_0^2|_{T_{i}}]$ with $i=2,\ldots, 6$.
Finally, for the last two intervals $i=7,8$ a 5-point Gaussian integration has been chosen.
These evaluations require the sampling of the integrand 
function, $\Delta\langle S_G\rangle^\infty_{\vec\xi}/\Delta\xi_k$,
at several values of the bare coupling, prescribed by the quadrature rules.
At each value of $g_0^2$ the integrand function is estimated with two independent 
numerical simulations of the pure SU$(3)$ Yang-Mills theory
at the same bare parameters and shifts $\vec\xi =(1\pm 2a/L_0,0,0)$.
\begin{figure}[h]
    \centering
    \includegraphics[width=0.4\textwidth]{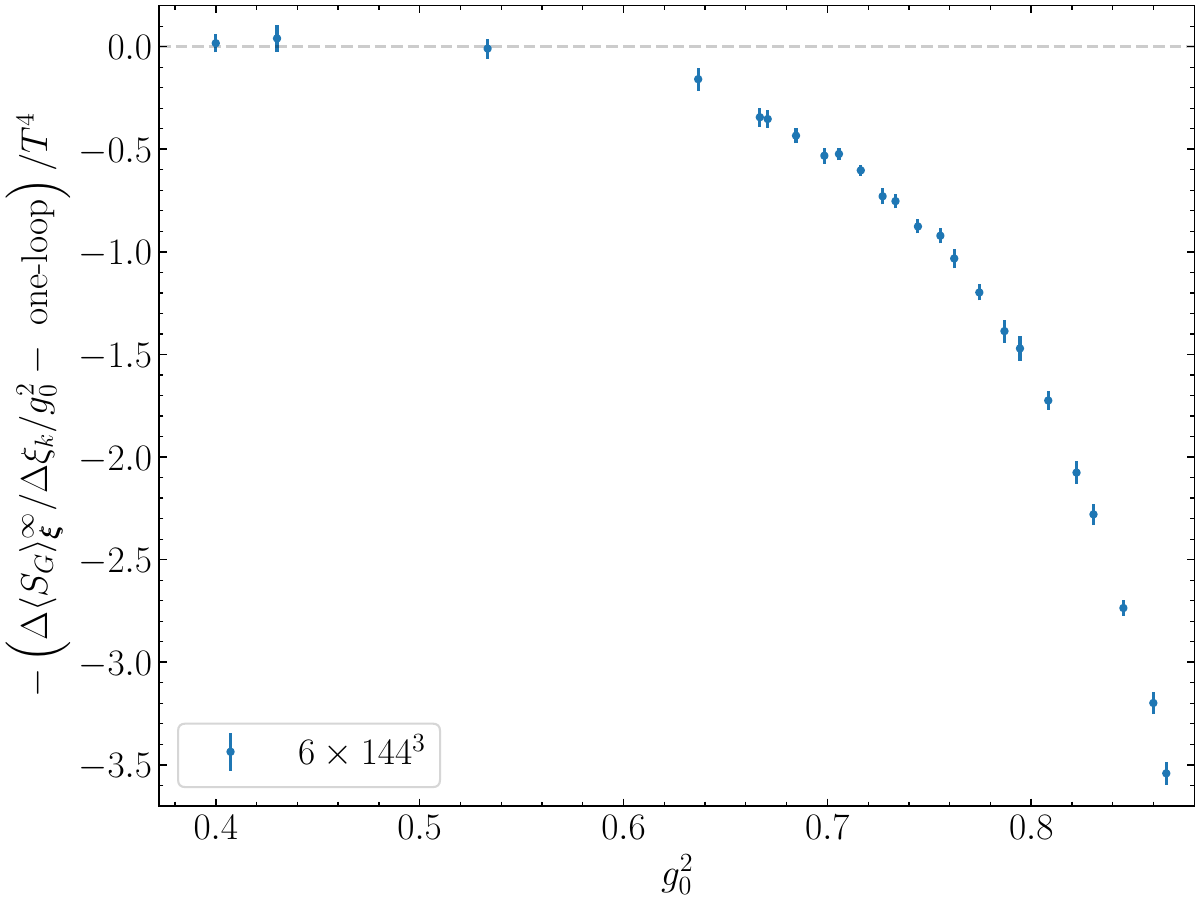}
    \caption{Derivative in the shift of the pure gauge action as a function 
    of $g_0^2$ for $L_0/a=6$. For convenience we subtracted from the data the 
    result at one-loop order in lattice perturbation theory.}
    \label{fig:g02_integral}
\end{figure}

Figure~\ref{fig:g02_integral} shows the resulting integrand for the $L_0/a=6$ case,
where its one-loop value has been subtracted. Notice that
at values of the squared coupling below $g_0^2=0.5$ the integrand is well compatible with
the perturbative prediction to which it can be safely connected. As we observed in Ref.~\cite{Giusti:2015daa},
the integrand is a smooth function and any potential systematic effect arising from the
choice of the 3-point or 5-point Gaussian quadrature methods between the $g_0^2$ values
of two successive temperatures, is negligible compared to the statistical uncertainties.\\[0.25cm]
\textbf{\textit{Appendix C: Monte Carlo results}} ---
In Table~\ref{tab:matel} we give the values of  
$\frac{\Delta f^\infty_{\vec\xi}}{\Delta\xi_k}$ and 
$\frac{\Delta (f_{\vec\xi}-f^\infty_{\vec\xi})}{\Delta\xi_k}$
for each lattice simulated together with the extension of the compact direction, $L_0/a$, and 
the inverse bare gauge coupling, $6/g_0^2$.
The computation of $\frac{\Delta f^\infty_{\vec\xi}}{\Delta\xi_k}$ is described in Appendix~B. 
We collected about 6000 pure gauge measurements per quadrature point for $L_0/a=4,6$ and around 40000 for $L_0/a=8,10$.
In Appendix~A we present the calculation of $\frac{\Delta (f_{\vec\xi}-f^\infty_{\vec\xi})}{\Delta\xi_k}$.
For $L_0/a=4,6$ we collected approximatively 200 measurements of the chiral condensate per Gaussian point. 
For $L_0/a=8,10$ the number of measurements per Gaussian point is about 400 in the first integration domain,
1000 in the second domain and 4000 in the third domain.
These numbers were tuned in order to minimize the cost of computing the full integral at the target statistical precision. To
this end, we took into account the weight of the individual ensembles in the integral, and the dependence of their variance,
autocorrelation, and computational cost with $m_q$.
The statistical errors have been estimated using the Gamma-method~\cite{Wolff:2003sm} in the implementation 
of Refs.~\cite{Joswig:2022qfe,Ramos:2018vgu}.\\[0.25cm]
\begin{table*}[t!]
\begin{center}
\setlength{\tabcolsep}{.30pc}
\begin{tabular}{|r|ccc|ccc|ccc|}
\hline
& & & & & & & & & \\[-0.25cm]
$\frac{L_0}{a}$   &
$6/g_0^2$ & $\frac{\Delta f^\infty_{\vec\xi}}{\Delta \xi_k}$\scalebox{0.7}{$\times 10^4$}&
           $\frac{\Delta(f_{\vec\xi}-f^\infty_{\vec\xi})}{\Delta \xi_k}$\scalebox{0.7}{$\times 10^4$}&
$6/g_0^2$ & $\frac{\Delta f^\infty_{\vec\xi}}{\Delta \xi_k}$\scalebox{0.7}{$\times 10^4$}&
           $\frac{\Delta(f_{\vec\xi}-f^\infty_{\vec\xi})}{\Delta \xi_k}$\scalebox{0.7}{$\times 10^4$}&
$6/g_0^2$ & $\frac{\Delta f^\infty_{\vec\xi}}{\Delta \xi_k}$\scalebox{0.7}{$\times 10^4$}&
           $\frac{\Delta(f_{\vec\xi}-f^\infty_{\vec\xi})}{\Delta \xi_k}$\scalebox{0.7}{$\times 10^4$}\\[0.125cm]
\hline
\hline
& \multicolumn{3}{c|}{$T_0$}
& \multicolumn{3}{c|}{$T_1$}
& \multicolumn{3}{c|}{$T_2$} \\
\hline
 4 &  8.7325  &  37.861(24)  &   111.19(4)  & 8.3033 & 37.558(24)  &  110.73(4)  & 7.9794 &  37.260(24)  &  110.29(5)   \\
 6 &  8.9950  &   6.819(24)  &    15.25(5)  & 8.5403 &  6.768(24)  &   15.15(3)  & 8.2170 &   6.720(24)  &   15.24(4)   \\
 8 &    -     &   -        &    -       & 8.7325 &   2.078(9)  &  4.370(24)  & 8.4044 &    2.062(9)  &  4.381(22)   \\
10 &    -     &   -        &    -       & 8.8727 &  0.856(11)  &  1.740(19)  & 8.5534 &   0.850(11)  &  1.731(19)   \\
\hline
& \multicolumn{3}{c|}{$T_3$}
& \multicolumn{3}{c|}{$T_4$}
& \multicolumn{3}{c|}{$T_5$} \\
\hline
 4 & 7.6713 & 36.890(24)  &  109.98(6) & 7.3534 &  36.381(24)  &  109.69(5) & 7.0250 & 35.625(25)  &  109.31(6)  \\
 6 & 7.9091 &  6.656(24)  &   15.14(4) & 7.5909 &   6.564(24)  &   15.15(4) & 7.2618 &  6.423(24)  &   15.16(4)  \\
 8 & 8.0929 &   2.041(9)  &  4.347(24) & 7.7723 &    2.010(9)  &  4.379(24) & 7.4424 &   1.964(9)  &  4.439(27)  \\
10 & 8.2485 &  0.843(11)  &  1.725(20) & 7.9322 &   0.830(11)  &  1.766(21) & 7.6042 &  0.813(11)  &  1.737(24)  \\
\hline
& \multicolumn{3}{c|}{$T_6$}
& \multicolumn{3}{c|}{$T_7$}
& \multicolumn{3}{c|}{$T_8$} \\
\hline
 4 & 6.7079 & 34.499(25)  &  109.30(6) & 6.3719 & 32.397(25)  &  109.92(9)  &    -   &    -      &     -       \\
 6 & 6.9433 &  6.201(24)  &   15.30(5) & 6.6050 &  5.767(24)  &   15.61(5)  & 6.2735 &  4.711(26)  &   16.43(7)    \\
 8 & 7.1254 &   1.890(9)  &  4.436(22) & 6.7915 &   1.748(9)  &    4.57(4)  & 6.4680 &   1.416(9)  &    4.89(4)    \\
10 & 7.2855 &  0.784(11)  &  1.793(24) & 6.9453 &  0.722(11)  &  1.839(20)  & 6.6096 &  0.571(11)  &  1.939(23)    \\
\hline
\end{tabular}
\caption{\label{tab:matel}
Results for $\frac{\Delta f^\infty_{\vec\xi}}{\Delta \xi_k}$ and
$\frac{\Delta (f_{\vec\xi}-f^\infty_{\vec\xi})}{\Delta \xi_k}$ obtained from Monte Carlo
simulations. For each lattice we also report the size of the compact direction, $L_0/a$,
and the value of the inverse bare gauge coupling, $6/g_0^2$. The full set of bare parameters
specifying the lines of constant physics is reported in Table~4 of Ref.~\cite{DallaBrida:2021ddx}}
\end{center}
\end{table*}
\begin{table}[th]
    \centering
    \begin{tabular}{|c|c|cc|}
    \hline
    $T$ & $L/a$ &  $\langle T_{01}^{G,\{6\}}\rangle_{\vec\xi}$\scalebox{0.8}{$\times 10^3$} & 
                   $\langle T_{01}^{F,\{6\}}\rangle_{\vec\xi}$\scalebox{0.8}{$\times 10^3$}\\
    \hline
    \multirow{2}{*}{$T_1$} & $144$ & $-0.541(3)$ & $-1.2164(11)$ \\
                           & $288$ & $-0.539(3)$ & $-1.2183(11)$ \\
    \hline
    \multirow{2}{*}{$T_8$} & $144$ & $-0.458(4)$ & $-1.1280(19)$ \\
                           & $288$ & $-0.455(4)$ & $-1.1317(20)$ \\
    \hline
    \end{tabular}
    \caption{Expectation values of the bare energy-momentum tensor matrix elements.
      We consider two values of the spatial volume for  $L_0/a=6$
      at $T_1$ and $T_8$.
    }
    \label{tab:fveffects}
\end{table}
\textbf{\textit{Appendix D: Finite volume effects}} ---
At high temperature finite volume effects in the entropy are exponentially suppressed for $L\to\infty$
with an exponent $\propto LT$ \cite{Giusti:2012yj,DallaBrida:2021ddx}.
Since in our simulations $10\lesssim LT\lesssim 25$, we expect these effects to be very small, well 
below the statistical uncertainties. We have explicitly checked this at the temperatures
$T_1$ and $T_8$ for $L_0/a=6$.\\
The determination of the entropy density using Eq.~\eqref{eq:sshift} on volumes much larger than $L/a=144$
would be computationally very demanding. For checking 
finite volume effects we employ the relation~\cite{Giusti:2012yj,DallaBrida:2020gux}
\begin{equation}
    \frac{s}{T^3} = -\frac{1+\vec\xi^2}{\xi_k}\frac{1}{T^4}\langle T_{0k}^{R,\{6\}}\rangle_{\vec\xi}\,,
    \label{eq:s_T0k6}
\end{equation}
where we define the sextet component of the
renormalized energy-momentum tensor as~\cite{DallaBrida:2020gux}
\begin{equation}
    T_{\mu\nu}^{R,\{6\}} = Z_G^{\{6\}}(g_0^2)\,T_{\mu\nu}^{G,\{6\}} + Z_F^{\{6\}}(g_0^2)\,T_{\mu\nu}^{F,\{6\}}\,.
    \label{eq:T6R}
\end{equation}
We computed the bare matrix elements of the energy-momentum tensor at two values of the spatial size $L/a=144$ 
and $L/a=288$, on $L_0/a=6$ lattices at the bare parameters of $T_1$, $T_8$ and shift $\vec\xi=(1,0,0)$.
The results are reported in Table~\ref{tab:fveffects}. No finite size effects are observed
in the bare matrix elements. The statistical precision of the bare matrix elements has been tuned so
that the relative error on the entropy from
Eq.~(\ref{eq:s_T0k6}), with the renormalization constants estimated in perturbation theory, is comparable with
the one obtained from the results in Table~\ref{tab:matel}.\\[0.25cm] 
\textbf{\textit{Appendix E: Perturbative improvement}} ---
At fixed parameters $L_0/a$ and $g_0$ we remove the lattice artifacts at one-loop order
in lattice perturbation theory through the replacement,
\begin{equation}
    s(L_0/a, g_0^2) \to s(L_0/a, g_0^2)\cdot \frac{s_0 + s_2\left(\frac{g}{2\pi}\right)^2}
    {s_0{\scriptstyle(L_0/a)} + s_2{\scriptstyle(L_0/a)}\left(\frac{g}{2\pi}\right)^2}\,,
\end{equation}
where $g^2=\bar{g}_{\rm SF}^2(1/L_0)$ and in the denominator we have
the one-loop coefficients in lattice perturbation theory in the thermodynamic limit.
\begin{table}[th]
    \centering
    \begin{tabular}{|c|ccccc|}
    \hline
    $L_0/a$ & $4$ & $6$ & $8$ & $10$ & $\infty$ \\
    \hline
    $s_0$ & $4.719$ & $3.438$ & $3.145$ & $3.059$ & $2.969$ \\
    $s_2$ &  $-23.465$ & $-13.085$ & $-9.955$ & $-9.089$ & $-8.438$ \\
    \hline
    \end{tabular}
    \caption{Coefficients for the one-loop improvement of the lattice entropy density.}
    \label{tab:entropy_lpt}
\end{table}
Their values for the relevant $L_0/a$ are listed in Table~\ref{tab:entropy_lpt},
together with the results in the continuum limit. After improving at one-loop order,
the leading discretization effects are expected to be parametrically of $O(g^3a^2)$, since all
$O(g^0a^n)$ and $O(g^2a^n)$ effects have been subtracted. On general grounds, at finite 
temperature we expect odd powers in the coupling to be present.\\[0.25cm]
\textbf{\textit{Appendix F: Continuum limit extrapolations}} --- 
We estimated the entropy density in the continuum
limit by performing a combined fit of the data
at the various temperatures at finite lattice spacing.
We have considered the following fit function
\begin{equation}
    s(T_i, a/L_0)/T_i^3 = c_i + d_2g_i^3 (a/L_0)^2+d_3g_i^3 (a/L_0)^3\,,
    \label{eq:fit_type}
\end{equation} 
where $i=0,\ldots,8$, $g_i\equiv\bar{g}_{\rm SF}(\sqrt{2}T_i)$, $c_i$ are the continuum values,
and $d_2,d_3$ parametrize the discretization effects.\\
We first considered fits to the data with $L_0/a=4,6,8,10$, where either $d_3=0$ or $d_3\neq0$. 
Both fits have $\chi^2/\chi^2_{\rm exp}\approx 1$, with the former giving 
values for $c_i$ which are statistically $3$-$4$ times more precise.
Although compatible within the
larger errors of the second fit, the extrapolated  central values of  the first fit are systematically higher.
We then compared the fit with $L_0/a\geq4$ and $d_3\neq0$ to the one with $L_0/a>4$ and $d_3=0$.
For the latter we get $\chi^2/\chi^2_{\rm exp}\approx 0.7$ and comparable errors for the $c_i$.
The two fits are in good agreement with no clear systematic difference. 
This hints to the fact that the data at $L_0/a=4$ are likely affected by discretization effects 
of higher order than $a^2$. 
To be conservative, we thus used the data with $L_0/a=4$ only to estimate the size of 
the $O(a^3)$ contributions and included these as a systematic error to the points at finer lattice spacing. 
More precisely, we used the value of $d_3$ determined above, and modified the weights of the fits by adding 
in quadrature to the statistical errors of the points a systematic effect $d_3g_i^3(a/L_0)^3$. 
The final best fit is the one considering data with $L_0/a>4$, $d_3=0$ in the fit ansatz Eq.~(\ref{eq:fit_type}), 
and the modified weights. The continuum results are stable against adding a term $\propto g_i^4 (a/L_0)^2$ to 
Eq.~(\ref{eq:fit_type}). To further corroborate the robustness of the fit we also checked the impact of logarithmic 
corrections to the $O(a^2)$ effects, modeled as $a^2[\bar g^2_{\rm SF}(\pi/a)]^\gamma$~\cite{Husung:2019ytz,Husung:2021mfl,Husung:2022kvi}. By varying the effective anomalous dimension $\gamma\in[-1,1]$ the results change by less than $1$ standard deviation.\\ 
As an additional test for the soundness of the procedure, we repeated the whole analysis replacing $d_3g_i^3(a/L_0)^3$
with $d_4g_i^3(a/L_0)^4$ in Eq.~(\ref{eq:fit_type}). We obtain perfectly compatible results for the $c_i$
with errors that are $10\%$-$20\%$ smaller. Similar conclusions hold when replacing $g_i^3\to g_i^4$ in both $a^2$ and
$a^3$ terms.\\
All statistical correlations have been properly taken into account in the final results using the
tools of Refs.~\cite{Joswig:2022qfe,Ramos:2018vgu} for the error propagation. In particular, we propagated the statistical 
uncertainties deriving from the definition of the lines of constant physics~\cite{DallaBrida:2018rfy,DallaBrida:2021ddx}, 
and the correlations introduced by the integration in the bare coupling. While the former have a negligible 
effect on the errors of the final $c_i$ values, neglecting the latter gives errors that would be
about $10$-$20\%$ smaller.



\end{document}